\documentclass[journal]{IEEEtran}
\usepackage{graphicx}
\usepackage{booktabs}
\usepackage{diagbox}
\usepackage{algorithmic}
\usepackage{amsmath}
\usepackage{amsfonts}
\usepackage{url}
\usepackage{cite}
\ifCLASSINFOpdf
\else
\fi
\begin{document}
\title{ReCoFlow-MRI: Few-Step Conditional Flow Matching for GNL-Induced Nonlinear MRI Reconstruction}
\author{Qidi Luo, Hongli Chen, Jialiang Zhang, Yuting He, Yang Gao,
Chunyi Liu, Yaohui Wang, Feng Liu, and Shanshan Shan%
\thanks{This work was supported in part by the National Natural
Science Foundation of China under Grants 62301352 and 62301616,
and in part by the Natural Science Foundation of Jiangsu Province
under Grant BK20255001.}%
\thanks{Qidi Luo and Hongli Chen contributed equally to this work.
(Corresponding author: Shanshan Shan.)}%
\thanks{Qidi Luo, Jialiang Zhang, Yuting He, and Shanshan Shan
are with the State Key Laboratory of Radiation Medicine and
Protection, School of Radiation Medicine and Protection,
Collaborative Innovation Center of Radiological Medicine
of Jiangsu Higher Education Institutions,
Soochow University, Suzhou 215123, China
(e-mail: 20244220002@\allowbreak stu.\allowbreak suda.\allowbreak edu.\allowbreak cn;
20244220038@\allowbreak stu.\allowbreak suda.\allowbreak edu.\allowbreak cn;
20254220047@\allowbreak stu.\allowbreak suda.\allowbreak edu.\allowbreak cn;
ssshan@\allowbreak suda.\allowbreak edu.\allowbreak cn).}%
\thanks{Shanshan Shan is also with the Biomedical Basic Research Center
(BBRC) of Jiangsu Province and the Jiangsu Key Laboratory
of Radiation Injury Prevention and Hazardous Substance
Control, Suzhou, China.}%
\thanks{Hongli Chen and Feng Liu are with the School of
Electrical Engineering and Computer Science,
The University of Queensland, Brisbane, QLD 4072, Australia
(e-mail: hongli.chen@uq.edu.au;
feng@eecs.uq.edu.au).}%
\thanks{Yang Gao is with the School of Computer Science
and Engineering, Central South University,
Changsha 410083, China
(e-mail: yang.gao@csu.edu.cn).}%
\thanks{Chunyi Liu is with the Medical School,
Nanjing University, Nanjing 210993, China
(e-mail: chunyi\_liu@nju.edu.cn).}%
\thanks{Yaohui Wang is with the Institute of Electrical Engineering, Chinese Academy of Science, Beijing 100190, China (e-mail: yhw@mail.iee.ac.cn).}%
}
\markboth{IEEE Transactions on BIOMEDICAL ENGINEERING}%
{Luo et al.: ReCoFlow-MRI}
\maketitle
\begin{abstract}
\textit{Objective:} Accelerated magnetic resonance imaging (MRI) reconstruction is challenging in the presence of gradient nonlinearity (GNL). GNL-induced nonlinear spatial encoding causes geometric distortions that compromise image-guided radiotherapy. Diffusion models require extensive sampling and rarely account for GNL. Flow matching shortens sampling, but velocity matching does not directly minimize the reconstruction error. We propose ReCoFlow-MRI, a conditional flow-matching framework for few-step GNL-corrected reconstruction. \textit{Methods:} The calibrated GNL displacement field is integrated into the nonlinear encoding model to construct a distortion-corrected zero-filled image, which conditions a velocity field learned along a linear path from Gaussian noise to fully sampled images. The pretrained network is fine-tuned with a pixel-supervised endpoint loss on its five-step reconstruction. \textit{Results:} On simulated fastMRI brain data, five-step ReCoFlow-MRI achieved the highest PSNR and SSIM among compared methods, including 128-step diffusion sampling. On MRI-Linac phantom and volunteer data, it achieved effective reconstruction and reduced the maximal geometric error from 24.95 to 2.12 mm, with 98.39\% of the 3718 fiducial markers below 1 mm. \textit{Conclusion:} ReCoFlow-MRI jointly addresses undersampling artifacts and geometric distortions within a few-step generative reconstruction. \textit{Significance:} It provides a potential solution for fast and geometrically accurate imaging in image-guided radiotherapy.

\end{abstract}

\IEEEpeerreviewmaketitle
\section{Introduction}
\IEEEPARstart{M}{agnetic} resonance imaging (MRI) provides non-ionizing imaging with high soft-tissue contrast and spatial resolution, which has been integrated with linear accelerators (Linac) for real-time image-guided radiotherapy~\cite{goodburn2022future}. Both acquisition efficiency and geometric fidelity are critical for accurate tumor treatments using such MRI-Linac systems. However, prolonged MRI acquisition times may cause motion artifacts and degrade real-time image guidance~\cite{zaitsev2015motion}. In addition, geometric distortions induced by gradient nonlinearity (GNL) restrict anatomical structure accuracy, potentially leading to target underdosage and unnecessary radiation on surrounding normal tissues~\cite{weygand2016spatial}. Undersampling k-space can substantially accelerate MRI acquisition, but causing aliasing artifacts. Meanwhile, the presence of GNL deviates the forward linear gradient encoding process~\cite{janke2004spherical,tao2015integrated}. Therefore, reconstructing high-fidelity images from undersampled k-space measurements while simultaneously correcting GNL-induced geometric distortions constitutes a challenging nonlinear MRI reconstruction problem~\cite{Shan2019GFD,Shan2020Geometric}.

Various methods have been developed for accelerated MRI reconstruction~\cite{wang2016accelerating,yang2017dagan,knoll2020deep,liang2020deep}. Compressed sensing (CS)~\cite{lustig2007sparse,candes2008introduction} exploits sparsity or other handcrafted priors to recover high-quality images from undersampled k-space measurements. However, CS-based methods generally use iterative optimization algorithms, which require large computational cost. Deep learning techniques such as U-Net-based architectures, subsequently enable data-driven MRI image reconstruction by learning nonlinear mappings from undersampled measurements to fully sampled images~\cite{heckel2024deep}. Model-based methods~\cite{zhang2018ista,aggarwal2018modl,shan2023distortion,chen2026hifimamba} incorporate MRI physical models into network architectures and apply convolutional neural network (CNN) modules for image regularization, which have demonstrated improved interpretability than data-driven methods~\cite{aggarwal2018modl,ramzi2022nc,zhang2018ista,hammernik2018variational,sriram2020end}. However, the performance of model-based networks will be degraded dramatically for highly undersampled MRI acquisitions~\cite{muckley2021results}.

More recently, diffusion-based generative models have been applied to accelerated MRI reconstruction by learning image priors from fully sampled MR images~\cite{chung2022score,song2021solving,cao2024high}. However, the iterative sampling procedures require repeated network evaluations, resulting in substantial computational cost. Early efforts to reduce this cost focus on modifying the sampling process of pretrained diffusion models~\cite{song2021ddim,chung2022come}, but the sample quality of denoising diffusion implicit models (DDIM) degrades sharply when few sampling steps are used~\cite{salimans2022progressive}. Subsequent work introduces progressive teacher--student distillation to transfer the behavior of a multi-step teacher sampler to a student model requiring fewer steps, at the cost of additional training~\cite{salimans2022progressive}. Flow matching provides a direct training objective for learning a velocity field along prescribed transport paths, allowing deterministic generation and faster sampling without requiring a pretrained diffusion teacher~\cite{lipman2022flow}. These generative approaches rarely account for GNL-induced perturbations of the spatial encoding. In addition, standard flow matching learns the velocity field through velocity matching, which does not directly minimize the reconstruction error after few-step integration.
\begin{figure*}[!t]
\centering
\includegraphics[width=0.95\textwidth]{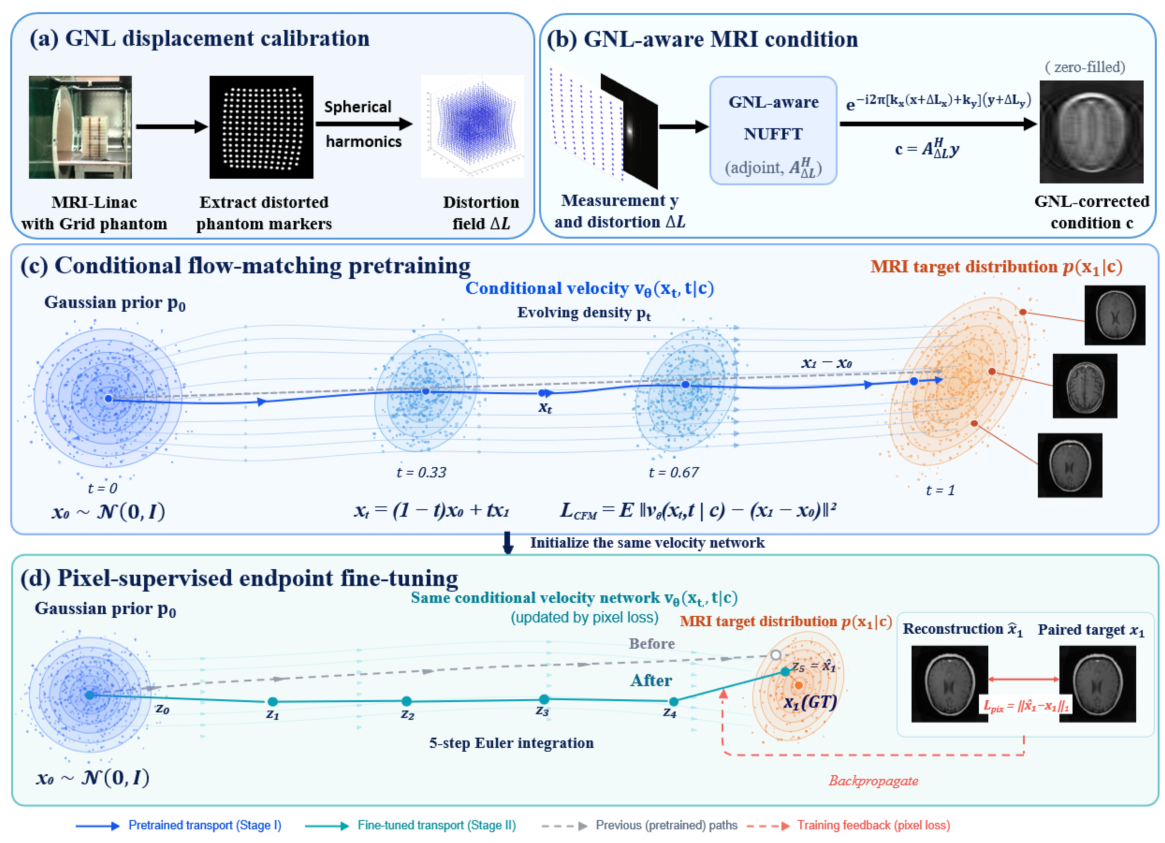}
\caption{Overview of the proposed ReCoFlow-MRI framework. 
(a) GNL displacement calibration. A grid phantom is scanned on the MRI-Linac system, and the distorted phantom markers are extracted to estimate the system-specific GNL displacement field \(\Delta L\) using spherical harmonic modeling.
(b) GNL-aware MRI condition. The calibrated displacement field is incorporated into the nonlinear spatial encoding model, and a GNL-aware NUFFT adjoint is applied to the undersampled measurement \(y\) to obtain the GNL-corrected zero-filled image \(c\), which serves as the conditioning image.
(c) Conditional flow-matching pretraining. Starting from Gaussian noise \(x_0\sim p_0\), the conditional velocity field \(v_\theta(x_t,t\mid c)\) is optimized to transport samples toward the fully sampled MRI target distribution \(p(x_1\mid c)\) along the prescribed interpolation path using the conditional flow-matching loss \(\mathcal{L}_{\mathrm{CFM}}\).
(d) Pixel-supervised endpoint fine-tuning. The pretrained velocity network is fine-tuned using an image-domain $\ell_1$ loss between the five-step Euler reconstruction \(\hat{x}_1\) and its paired target \(x_1\). Gradients are propagated only through the final Euler step, and the network parameters are shared across all integration steps.}
\label{fig:overview}
\end{figure*}

In this work, we propose ReCoFlow-MRI, a few-step conditional flow-matching method for GNL-induced nonlinear MRI reconstruction. The calibrated GNL displacement field is incorporated into the MRI encoding model to construct a zero-filled conditioning image that provides measurement-specific anatomical information in the corrected image coordinates. Conditioned on this image, a velocity network learns transport from Gaussian noise to fully sampled targets through flow-matching pretraining, enabling reconstruction by few-step numerical integration. We further fine-tune the pretrained network with a pixel-supervised endpoint loss to directly supervise the reconstructed image against its paired target. The contributions of this work are as follows:
\begin{enumerate}
\item We propose ReCoFlow-MRI, a conditional flow-matching framework that jointly addresses accelerated MRI reconstruction and GNL correction. The framework incorporates calibrated GNL encoding information into a measurement-specific conditioning image to guide reconstruction.
\item We introduce pixel-supervised endpoint fine-tuning to complement conditional velocity matching. This training strategy directly supervises the reconstructed image produced by few-step integration, improving reconstruction fidelity.
\item We demonstrate accelerated MRI reconstruction and geometric distortion correction with five Euler steps on fastMRI, MRI-Linac acquired phantom and volunteer data. Comparative experiments and ablation studies evaluate reconstruction quality and the contributions of the proposed conditioning and endpoint fine-tuning.
\end{enumerate}

\section{Related Work}
\subsection{Conventional GNL Correction Methods}
Conventional GNL correction is commonly performed after image reconstruction, which maps distorted image coordinates back to corrected spatial locations given pre-characterized gradient field information~\cite{janke2004spherical,doran2005complete,Jovicich2006}. Image-domain interpolation is normally utilized to estimate the coordinate mapping information, and has been implemented on most commercial MR scanners. However, the image domain-based method will introduce signal blurring and resolution loss because of the intrinsic smoothing effect of the interpolation operation. This is particularly problematic at the edges of a large field of view (FOV)~\cite{tao2015integrated}.

\subsection{Nonlinear MRI Reconstruction and Existing Algorithms}
To overcome the limitations of image domain-based methods, subsequent studies incorporate GNL information directly into the MR spatial encoding model and reconstruct distortion-corrected images from acquired k-space data~\cite{tao2015integrated}. In this strategy, GNL induces a nonlinear spatial gradient encoding process, which is typically modeled by using nonuniform fast Fourier transform (NUFFT) operators. Thus, the GNL-corrected images can be recovered by solving a nonlinear MRI reconstruction problem~\cite{tao2015integrated}.

Tao et al.~\cite{tao2015integrated,tao2015noncartesian} apply CS-based approach to integrate GNL information into Cartesian and non-Cartesian gradient encoding models and reconstruct GNL-corrected MR images using optimization algorithms. Model-based deep learning methods, e.g., DCReconNet~\cite{shan2023distortion} and RebinNet~\cite{Shan2024} incorporate nonlinear gradient encoding model into an interpretable unrolled architecture for GNL correction. Meanwhile, effective regularization modules such as CNN are trained to suppress image artifacts caused by k-space undersampling.

\subsection{Diffusion-Based MRI Reconstruction}
Diffusion models have been applied to accelerated MRI reconstruction by combining learned image priors with measurement constraints during iterative sampling~\cite{huang2024noise,chu2025highly}. DiffuseRecon adopts a denoising diffusion probabilistic model~\cite{ho2020denoising} and incorporates acquired k-space data into the reverse denoising process~\cite{peng2022towards}. Jalal et al. combines a learned image-prior score with a measurement-likelihood gradient through annealed Langevin dynamics for multicoil reconstruction~\cite{jalal2021robust}. Score-MRI uses a continuous-time score model and interleaved reverse stochastic differential equation (SDE) sampling with data-consistency updates~\cite{chung2022score,song2021sde}. These iterative reconstruction procedures require repeated network evaluations and measurement-operator computations.

Efforts to reduce sampling cost have addressed both sampling updates and initialization. DDIM permits sampling over a reduced sequence of time steps without retraining the denoising network, while DPM-Solver reduces network evaluations using a dedicated high-order solver for diffusion ordinary differential equations (ODEs)~\cite{song2021ddim,lu2022dpmsolver}. For inverse problems including MRI reconstruction, Come-Closer-Diffuse-Faster (CCDF) applies forward diffusion to an initial reconstruction and begins reverse sampling at an intermediate noise level, reducing the number of reverse sampling steps~\cite{chung2022come}.

Distillation reduces sampling steps through additional teacher--student training. Progressive distillation (PD) trains a student to reproduce two teacher sampling steps in one update; successive distillation stages halve the required sampling steps~\cite{salimans2022progressive}. For example, SSDM-MRI combines iterative selective distillation of a pretrained conditional diffusion model with shortcut reverse sampling. Sampling starts from a forward-diffused zero-filled reconstruction, and the distilled model reconstructs the image with a single reverse sampling step~\cite{liu2026highly}.

\subsection{Flow Matching}
Flow matching learns a velocity field using supervision constructed from data and prescribed probability paths, which does not require a pretrained teacher to generate targets. Lipman et al. formulate flow matching as vector-field regression and introduce a conditional objective that permits training without numerically integrating the transport ODE~\cite{lipman2022flow}. Liu et al. propose rectified flow, which learns an ODE that follows straight paths between samples of two distributions~\cite{liu2022flow}. Tong et al. generalize conditional flow matching into different couplings between source and target samples and introduce minibatch optimal transport to construct training pairs~\cite{tong2024improving}. Zhang et al. proposed multi-modal straight flow matching, which learns transport from undersampled images to fully sampled targets and uses complementary MR modalities to guide reconstruction~\cite{zhang2024mutli}. At inference, samples are generated by numerically integrating the learned velocity field.

Although flow matching has been investigated for accelerated MRI reconstruction, its application to GNL-induced nonlinear spatial encoding has not been sufficiently explored. In addition, standard flow matching learns a velocity field through velocity matching, which does not directly minimize the reconstruction error after few-step integration. It forms a mismatch between fitting the velocity field along training paths and minimizing the reconstruction error against the paired fully sampled target.
\section{Methods}
\subsection{Nonlinear MRI Reconstruction Model}
Let $x$ denote the target image on the nominal spatial grid.
GNL displaces the actual encoding location from its nominal position $\mathbf{L}=[L_x,L_y,L_z]^{\mathsf T}$ by a spatially varying displacement field $\Delta L=[\Delta L_x,\Delta L_y,\Delta L_z]^{\mathsf T}$.
For two-dimensional MRI reconstruction, only the in-plane
displacement components are incorporated into the nonlinear
spatial encoding model.
For a k-space coordinate $\mathbf{k}=[k_x,k_y]^{\mathsf T}$,
the nominal Fourier encoding matrix and the GNL-induced
phase modulation matrix are respectively defined as~\cite{shan2023distortion,tao2015integrated,tao2015noncartesian}
\begin{equation}
[\mathbf{F}]_{\mathbf{k},\mathbf{L}}
=\exp[-i2\pi(k_xL_x+k_yL_y)],
\end{equation}
\begin{equation}
[\mathbf{E}_{\Delta L}]_{\mathbf{k},\mathbf{L}}
=\exp[-i2\pi(k_x\Delta L_x+k_y\Delta L_y)].
\end{equation}
The GNL-aware encoding operator is defined as $\mathbf{A}_{\Delta L}=\mathbf{E}_{\Delta L}\odot\mathbf{F}$, where $\odot$ denotes element-wise multiplication. The undersampled k-space measurement model is
\begin{equation}
y=\mathbf{M}\mathbf{A}_{\Delta L}x,
\label{eq:gnl_measurement}
\end{equation}
where $\mathbf{M}$ denotes the undersampling mask, and $y$ is the resulting masked k-space data, in which unacquired entries are set to zeros. When $\Delta L=\mathbf{0}$, $\mathbf{A}_{\Delta L}$ becomes the nominal Fourier encoding operator. For Cartesian k-space, the GNL-aware encoding operator is implemented using a type-I NUFFT. The GNL-corrected images can be obtained by solving the nonlinear MRI reconstruction problem defined in Eq. (3).

\subsection{GNL Field Calibration and Conditional Image Construction}
As illustrated in Fig.~\ref{fig:overview}(a), the GNL field is calibrated using a water-filled grid phantom~\cite{Shan2020Geometric,Shan2021ReUINet} containing fiducial markers at known nominal locations. Let $\mathbf{L}_{q}^{\mathrm{nom}}$ and $\mathbf{L}_{q}^{\mathrm{meas}}$ denote the nominal and MRI-measured positions of the $q$th matched marker, respectively. The discrete displacement measured at that location is
\begin{equation}
\boldsymbol{\delta}_{q}
=\mathbf{L}_{q}^{\mathrm{meas}}-\mathbf{L}_{q}^{\mathrm{nom}},
\qquad q=1,\ldots,Q,
\end{equation}
where $Q$ is the number of matched markers used for calibration. The displacement field is represented by a spherical harmonic expansion~\cite{janke2004spherical,Weavers2017,tao2017calibration}:
\begin{equation}
\Delta L(r,\vartheta,\varphi)
=\sum_{\ell=0}^{P}\sum_{m=0}^{\ell}
\mathbf{a}_{\ell m}r^\ell
Y_\ell^m(\vartheta,\varphi),
\end{equation}
where \((r,\vartheta,\varphi)\) denote the spherical coordinates, \(Y_\ell^m\) represents the spherical harmonic basis function, \(\mathbf a_{\ell m}\) denotes the corresponding vector of spherical harmonic coefficients for the three spatial directions, and \(P\) is the expansion order. The coefficients are estimated from the measured marker displacements using the Gauss-Newton iteration algorithm~\cite{Shan2020Geometric}. As shown in Fig.~\ref{fig:overview}(b), based on the calculated \(\Delta L\), the GNL-corrected zero-filled image is then obtained by applying the adjoint of the GNL-aware encoding operator:
\begin{equation}
c \equiv x_{\mathrm{zf}}^{\mathrm{correct}}
= \mathbf{A}_{\Delta L}^{\mathsf H} y,
\end{equation}
where $\mathbf{A}_{\Delta L}^{\mathsf H}$ is implemented using the adjoint NUFFT~\cite{Fessler2003}. The GNL-aware adjoint NUFFT maps the undersampled k-space measurements onto the nominal image grid using the calibrated displacement field. The resulting image $c$ conditions the velocity predictions in the proposed network.

\subsection{Conditional Flow-Matching Transport}
The GNL-corrected reconstruction can be formulated as transport from a Gaussian noise distribution to $p(x_1\mid c)$, the distribution of distortion-corrected images conditioned on $c$, as shown in Fig.~\ref{fig:overview}(c).
\subsubsection{Linear Conditional Transport Path}
Let $\mathcal{D}_{y}=\{(y_i,x_{1,i})\}_{i=1}^{N}$ denote $N$ paired measurements and fully sampled, distortion-corrected targets, where $x_1$ corresponds to the image $x$ in Eq.~\eqref{eq:gnl_measurement}. The associated target--condition pairs form $\mathcal{D}=\{(x_{1,i},c_i)\}_{i=1}^{N}$, where $c_i=\mathbf{A}_{\Delta L}^{\mathsf H}y_i$. Each target is regarded as a sample of $p(x_1\mid c)$ at its paired condition. For each pair $(x_1,c)$ sampled from $\mathcal{D}$, we independently sample $x_0$ and define the linear interpolation path~\cite{lipman2022flow,liu2022flow}
\begin{equation}
x_t=(1-t)x_0+tx_1,\quad x_0\sim\mathcal{N}(0,I),\; x_1\sim p(x_1\mid c),
\end{equation}
where $t\in[0,1]$.
\subsubsection{Conditional Velocity Field}
For each noise--target pair $(x_0,x_1)$, $x_t$ moves along the linear path with the constant velocity $u=dx_t/dt=x_1-x_0$, which depends on the unknown target $x_1$. The conditional velocity field is the expectation of $u$ over all noise--target pairs whose path passes through $z$ at time $t$ under condition $c$:
\begin{equation}
v(z,t\mid c)=\mathbb{E}[u\mid x_t=z,c].
\label{eq:cond_velocity}
\end{equation}
This field does not depend on $x_1$. The solution of
\begin{equation}
\begin{aligned}
\frac{d z(t)}{d t}&=v(z(t),t\mid c),\\
z(0)&=x_0,\qquad t\in[0,1],
\end{aligned}
\label{eq:flow_ode}
\end{equation}
has the same distribution as $x_t$ at every $t$~\cite{lipman2022flow,liu2022flow}, so its value at $t=1$ is a sample of $p(x_1\mid c)$. A network $v_{\theta}(z,t\mid c)$ with parameters $\theta$ approximates $v$ and takes $z$ and $c$, both defined on the nominal image grid, as channel-wise concatenated input.

\begin{figure}[!t]
\centering
\begin{minipage}{0.98\columnwidth}
\hrule
\vspace{3pt}
\noindent\textbf{Algorithm 1: Training and Inference of ReCoFlow-MRI}
\vspace{2pt}
\hrule
\vspace{3pt}
\footnotesize
\begin{algorithmic}[1]
\REQUIRE Paired training data $\mathcal{D}_{y}$,
$\mathbf{A}_{\Delta L}$, $v_{\theta}$, and $K=5$.
\ENSURE Trained parameters $\theta$; reconstruction $\widehat{x}_{1}$.

\STATE \textbf{Stage I: Conditional flow-matching pretraining}
\FOR{each minibatch $(y,x_1)$ sampled from $\mathcal{D}_{y}$}
    \STATE Construct $c=\mathbf{A}_{\Delta L}^{\mathsf H}y$
    using the GNL-aware NUFFT.
    \STATE Sample $x_0\sim\mathcal{N}(0,I)$
    and $t\sim\mathcal{U}(0,1)$.
    \STATE Set $x_t=(1-t)x_0+tx_1$ and $u=x_1-x_0$.
    \STATE $\mathcal{L}_{\mathrm{CFM}}
    \leftarrow \|v_{\theta}(x_t,t\mid c)-u\|_2^2$.
    \STATE Update $\theta$ using $\mathcal{L}_{\mathrm{CFM}}$.
\ENDFOR
\STATE \textbf{Stage II: Pixel-supervised endpoint fine-tuning}
\STATE Initialize $\theta$ with the pretrained parameters
from Stage I.
\FOR{each minibatch $(y,x_1)$ sampled from $\mathcal{D}_{y}$}
    \STATE Construct $c=\mathbf{A}_{\Delta L}^{\mathsf H}y$.
    \STATE Sample $x_0\sim\mathcal{N}(0,I)$.
    \STATE Set $z_0=x_0$ and $\Delta t=1/K$.
    \FOR{$k=0,\ldots,K-2$}
        \STATE Set $t_k=k\Delta t$.
        \STATE $z_{k+1}\leftarrow
        \operatorname{StopGrad}\left(
        z_k+\Delta t\,v_{\theta}(z_k,t_k\mid c)
        \right)$.
    \ENDFOR
    \STATE Set $t_{K-1}=(K-1)\Delta t$.
    \STATE $z_K=z_{K-1}+\Delta t\,
    v_{\theta}(z_{K-1},t_{K-1}\mid c)$.
    \STATE $\mathcal{L}_{\mathrm{pix}}
    \leftarrow \|z_K-x_1\|_1$.
    \STATE Update $\theta$ using $\mathcal{L}_{\mathrm{pix}}$.
\ENDFOR

\STATE \textbf{Inference for an undersampled measurement $y$}
\STATE Set $c=\mathbf{A}_{\Delta L}^{\mathsf H}y$,
$z_0\sim\mathcal{N}(0,I)$, and $\Delta t=1/K$.
\FOR{$k=0,\ldots,K-1$}
    \STATE $t_k=k\Delta t$.
    \STATE $z_{k+1}=z_k+
    \Delta t\,v_{\theta}(z_k,t_k\mid c)$.
\ENDFOR
\STATE \textbf{return} $\widehat{x}_1=z_K$.
\end{algorithmic}
\vspace{3pt}
\hrule
\end{minipage}
\end{figure}

\subsection{Training Objectives}
\label{sec:training_objectives}
Both training stages update the same network $v_\theta$ with the condition $c$ from Eq.~(6): Stage I fits the velocity field in Eq.~\eqref{eq:cond_velocity}, and Stage II supervises the image reconstructed by few-step integration of $v_\theta$.
\subsubsection{Conditional Flow-Matching Loss}
As illustrated in Fig.~\ref{fig:overview}(c), the conditional velocity field is pretrained using
\begin{equation}
\mathcal{L}_{\mathrm{CFM}}
=
\mathbb{E}_{(x_1,c)\sim\mathcal{D},x_0,t}
\left[
\left\|
v_\theta(x_t,t\mid c)-(x_1-x_0)
\right\|_2^2
\right].
\end{equation}
The expectation is taken over paired training targets and conditions $(x_1,c)$, Gaussian noise samples $x_0\sim\mathcal{N}(0,I)$, and times $t\sim\mathcal{U}(0,1)$. The minimizer of this regression over functions of $(x_t,t,c)$ is the conditional expectation in Eq.~\eqref{eq:cond_velocity}, so $v_\theta$ is trained to approximate $v$.

\subsubsection{Pixel-Supervised Endpoint Loss}
Velocity matching with $\mathcal{L}_{\mathrm{CFM}}$ supervises predictions at sampled states along the training interpolation paths. To supervise the reconstructed image produced by finite-step integration, an image-domain $\ell_1$ loss is applied to fine-tune the same network, as shown in Fig.~\ref{fig:overview}(d). The reconstructed image is obtained by integrating Eq.~\eqref{eq:flow_ode} with $v_\theta$ in place of $v$. With $K$ Euler steps, $\Delta t=1/K$, $t_k=k\Delta t$, and $z_0=x_0$, the updates are
\begin{equation}
\begin{aligned}
z_{k+1}&=z_k+\Delta t\,v_{\theta}(z_k,t_k\mid c),\\
&\qquad k=0,\ldots,K-1.
\end{aligned}
\label{eq:euler_update}
\end{equation}
To restrict backpropagation to the final step, we detach the state $z_{K-1}$ produced by the current network in the first $K-1$ steps and compute
\begin{equation}
\begin{aligned}
\bar{z}_{K-1}&=\operatorname{StopGrad}(z_{K-1}),\\
\widehat{x}_{1}&=\bar{z}_{K-1}+\Delta t\,
 v_{\theta}(\bar{z}_{K-1},t_{K-1}\mid c).
\end{aligned}
\end{equation}
The operator $\operatorname{StopGrad}$ preserves its input value and blocks gradient propagation through that input.
The endpoint loss compares this reconstruction with its paired fully sampled target:
\begin{equation}
\mathcal{L}_{\mathrm{pix}}
=\mathbb{E}_{\substack{(x_1,c)\sim\mathcal{D},\\
x_0\sim\mathcal{N}(0,I)}}
\left[\|\widehat{x}_1-x_1\|_1\right].
\end{equation}
Stage II updates the shared parameters $\theta$ using $\mathcal{L}_{\mathrm{pix}}$.

\subsection{Few-Step ODE Inference}
\label{sec:ode_inference}
At inference, the conditioning image $c$ is constructed from the undersampled measurement using Eq.~(6) and remains fixed throughout integration. Starting from $x_0\sim\mathcal{N}(0,I)$, the fine-tuned network is applied for the $K$ Euler steps in Eq.~\eqref{eq:euler_update}, and the reconstructed image is the terminal state $\widehat{x}_1=z_K$. Each reconstruction therefore requires $K$ evaluations of $v_\theta$. Algorithm~1 summarizes the complete training and inference procedure.

\begin{figure*}[!t]
\centering
\includegraphics[width=0.92\textwidth]{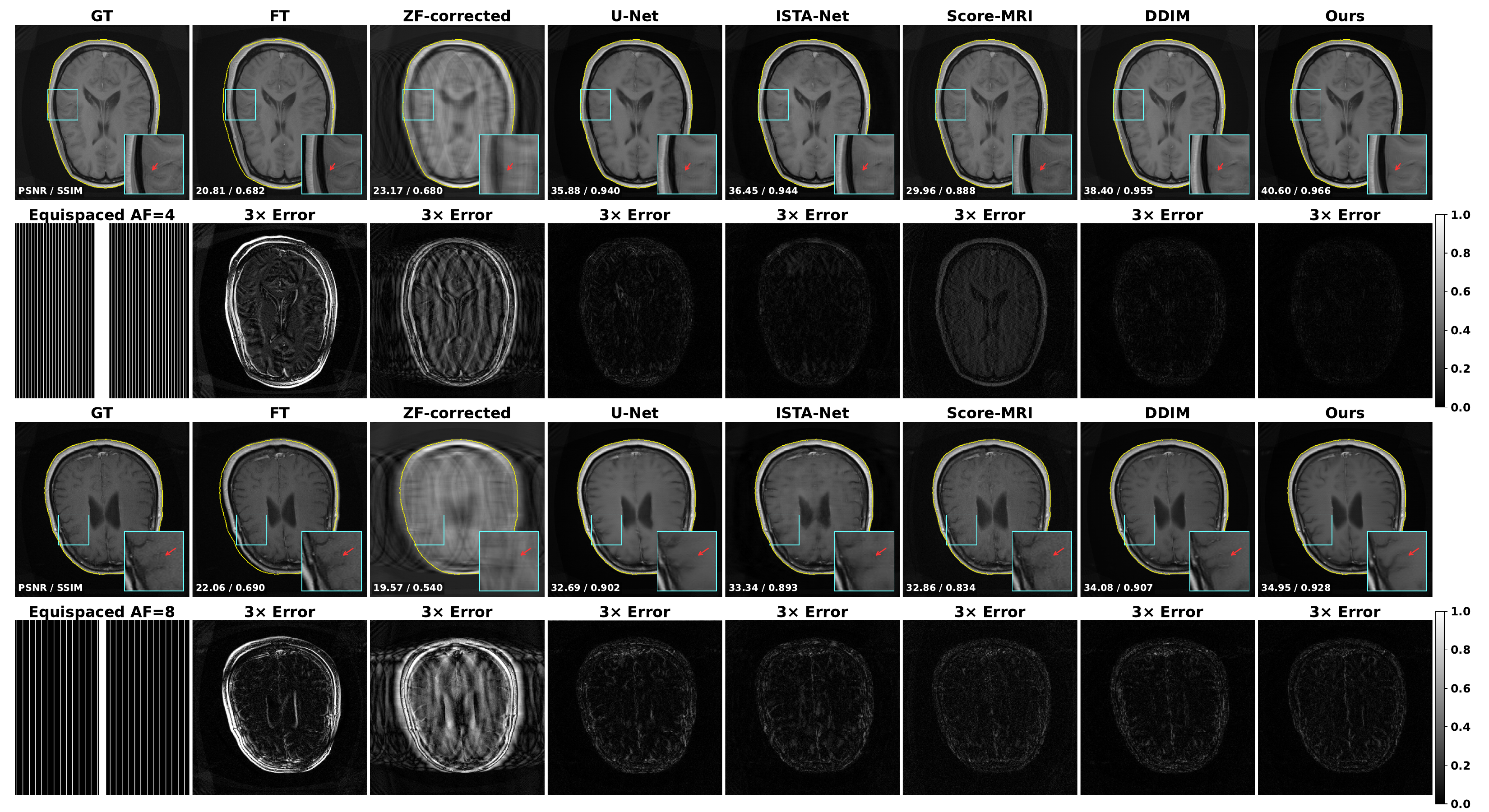}
\caption{Results on the fastMRI brain dataset under equispaced Cartesian undersampling with acceleration factors of $4\times$ and $8\times$. From left to right, it shows the fully sampled GT and FT reconstruction, followed by the ZF-corrected, U-Net, ISTA-Net, Score-MRI, DDIM, and the proposed ReCoFlow-MRI reconstructions with sampling masks below. Yellow lines denote the contour of GT brain images. Zoom-in regions highlight local anatomical details and PSNR/SSIM values are reported. The error maps are displayed at $3\times$ intensity.}
\label{fig:fastmri_equispaced}
\end{figure*}

\begin{table*}[!t]
    \centering
    \caption{Quantitative comparison on the fastMRI test set. PSNR is reported in dB; $\uparrow$ indicates that higher values are better. Best results are shown in bold.}
    \label{tab:comparison}
    \small
    \renewcommand{\arraystretch}{1.18}
    \setlength{\tabcolsep}{5.5pt}
    \begin{tabular}{lcccccccc}
    \toprule
    & \multicolumn{4}{c}{Equispaced Cartesian} & \multicolumn{4}{c}{Poisson-disc} \\
    \cmidrule(lr){2-5}\cmidrule(lr){6-9}
    & \multicolumn{2}{c}{E4$\times$} & \multicolumn{2}{c}{E8$\times$}
    & \multicolumn{2}{c}{P12$\times$} & \multicolumn{2}{c}{P15$\times$} \\
    \cmidrule(lr){2-3}\cmidrule(lr){4-5}\cmidrule(lr){6-7}\cmidrule(lr){8-9}
    Method
    & PSNR$\uparrow$ & SSIM$\uparrow$
    & PSNR$\uparrow$ & SSIM$\uparrow$
    & PSNR$\uparrow$ & SSIM$\uparrow$
    & PSNR$\uparrow$ & SSIM$\uparrow$ \\
    \midrule
    U-Net     & 34.88 & 0.939 & 30.73 & 0.896 & 32.53 & 0.904 & 32.13 & 0.899 \\
    ISTA-Net  & 35.05 & 0.944 & 30.86 & 0.889 & 35.35 & 0.920 & 34.54 & 0.911 \\
    Score-MRI & 32.88 & 0.887 & 30.67 & 0.830 & 32.06 & 0.847 & 31.82 & 0.839 \\
    DDIM      & 36.84 & 0.950 & 32.63 & 0.907 & 33.30 & 0.903 & 32.85 & 0.898 \\
    PD (8 steps) & 36.76 & 0.950 & 31.58 & 0.897 & 32.64 & 0.897 & 32.33 & 0.895 \\
    \midrule
    ReCoFlow-MRI (1 step) & 38.05 & 0.948 & 34.16 & 0.915 & 35.11 & 0.917 & 34.84 & 0.915 \\
    \textbf{ReCoFlow-MRI (5 steps)}
              & \textbf{39.17} & \textbf{0.961}
              & \textbf{34.85} & \textbf{0.930}
              & \textbf{35.78} & \textbf{0.928}
              & \textbf{35.17} & \textbf{0.922} \\
    \bottomrule
    \end{tabular}
\end{table*}

\section{Experiments and Results}
\subsection{Experimental Setup}
\subsubsection{Training Data Preparation and Network Training}
The public fastMRI T1POST brain dataset~\cite{zbontar2018fastMRI} was used for model training in this work. The brain MRI data were acquired at NYU Langone Health using 1.5 T and 3 T Siemens MRI scanners with axial T1-weighted post-contrast sequences. 8351 and 1068 brain images were used for model training and testing, respectively. All brain slices were resized into the same image size of $320\times320$. The GNL displacement field $\Delta L$ in the FOV of $30\times30\times30$ cm on an open-bore MRI-Linac was measured and calculated by our previously developed spherical harmonic model~\cite{shan2023distortion}. Based on the GNL field information, the GNL-corrupted undersampled k-space data was simulated for model training using the NUFFT operator in Eq.~\eqref{eq:gnl_measurement}. The 1D equispaced Cartesian mask along the phase encoding direction with acceleration factors (AFs) of $4\times$ and $8\times$, and 2D Poisson-disc masks with AFs of $12\times$ and $15\times$ were used in the proposed network.

The proposed ReCoFlow-MRI model was pretrained on an NVIDIA H100 GPU (80 GB) using the Adam optimizer with a learning rate of \(1\times10^{-4}\) and a weight decay of \(1\times10^{-6}\) for 180 epochs. The batch size was 10, and the learning rate was adjusted using a cosine annealing schedule. Endpoint fine-tuning and inference used $K=5$ Euler steps. The proposed models were implemented in PyTorch, and the source code will be made publicly available at https://github.com/shanshanshan3-hub/ReCoFlow. Human studies were approved by the Institutional Review Board (IRB).

\subsubsection{Evaluation Data Acquisition}
To evaluate the reconstruction performance of the proposed network, a 3D grid phantom, a brain phantom and volunteer brain were scanned on the open-bore Australian MRI-Linac scanner. The fully sampled k-space data acquired from the scanner were retrospectively undersampled using $12\times$ and $15\times$ Poisson-disc masks.

\paragraph{Grid phantom data}
A grid phantom with known 3718 fiducial markers was scanned using the body coil with a turbo spin echo (TSE) sequence. The acquisition matrix was $130\times110\times192$, with a spatial resolution of $1.8\times2.0\times1.8$~mm, an image bandwidth of 202~Hz, TE/TR of 15~ms/5.1~s, and R/L phase encoding direction.Two acquisitions were performed, with the phantom positioned at the scanner isocenter and at an off-center location, respectively to cover a large FOV. The fully sampled k-space data were retrospectively undersampled using $4\times$ and $8\times$ 1D equispaced Cartesian masks.

\paragraph{Brain phantom data}
A brain phantom was scanned using a six-channel head coil with a TSE sequence. The acquisition matrix was $256\times256\times40$, with a spatial resolution of $0.98\times0.98\times5$~mm, an image bandwidth of 203~Hz, TE/TR of 101~ms/12~s, and R/L phase encoding direction. The acquired k-space data were retrospectively undersampled using $12\times$ and $15\times$ 2D Poisson-disc masks.
\begin{figure*}[!t]
\centering
\includegraphics[width=0.92\textwidth]{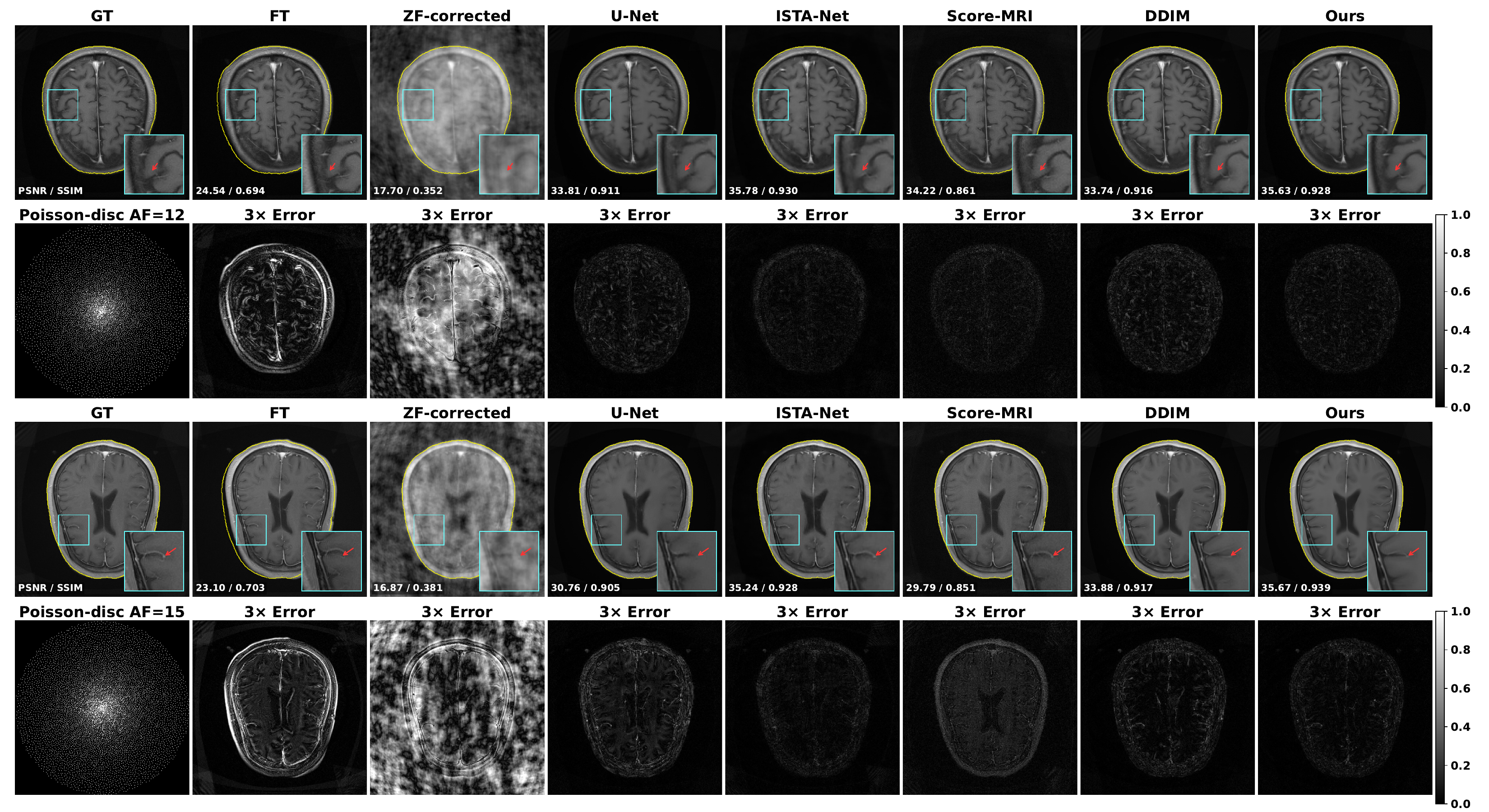}
\caption{Results on the fastMRI brain dataset under Poisson-disc undersampling with acceleration factors of 12× and 15×. The first and third rows show the GT, FT, ZF-corrected, U-Net, ISTA-Net, Score-MRI, DDIM, and the proposed ReCoFlow-MRI reconstructions, respectively. The second and fourth rows present zoomed-in regions indicated by cyan boxes and the corresponding error maps. Yellow contours are extracted from the GT images and overlaid on all reconstructions. Red arrows highlight local anatomical details. PSNR/SSIM values are reported, and the error maps are displayed at 3× intensity.}
\label{fig:fastmri_poisson}
\end{figure*}
\paragraph{Volunteer brain data}
A healthy volunteer's brain was scanned using a six-channel head coil. To cover the whole diameter of spherical volume (DSV), two scans were performed where the volunteer's brain was firstly centered at the scanner isocenter and then shifted by 5~cm along the $x$ direction as a second scan. The acquisition matrix was $256\times256\times12$, with a spatial resolution of $0.98\times0.98\times5$~mm, an image bandwidth of 203~Hz, TE/TR of 77~ms/8~s, and R/L phase encoding direction. The fully sampled k-space data were retrospectively undersampled using $12\times$ and $15\times$ 2D Poisson-disc masks.
\subsubsection{Evaluation Methods and Metrics}
In this study, U-Net~\cite{ronneberger2015unet}, ISTA-Net~\cite{zhang2018ista}, Score-MRI~\cite{chung2022score}, DDIM~\cite{song2021ddim} and PD~\cite{salimans2022progressive} were implemented to compare the reconstruction performance with the proposed method. For experimentally acquired grid phantom, brain phantom and volunteer brain data, ground truth (GT) images were obtained by applying optimization-based GNL correction to the fully sampled images. Standard inverse Fourier reconstruction was applied to fully sampled k-space data, yielding geometrically distorted images referred to as FT. The GNL-aware adjoint NUFFT was applied into undersampled k-space data to generate GNL-corrected zero-filled images, referred to as ZF-corrected. The U-Net was trained to learn the mappings between ZF-corrected images and target images. NUFFT operator was integrated into the ISTA-Net architecture for distortion-corrected reconstruction. For the Score-MRI implementation, NUFFT was applied to the data consistency module in the inference step. For the DDIM baseline, the GNL-corrected zero-filled image was concatenated with the noisy image as an additional input to the denoising network, and reconstruction was performed using 128 sampling steps. For PD, a conditional diffusion model using 128-step DDIM~\cite{song2021ddim} sampling was employed as the initial teacher, with the GNL-corrected zero-filled image concatenated with the network input as an additional condition. PD was then performed to successively reduce the sampling steps from 128 to 64, 32, 16, and 8, while maintaining the same conditioning strategy throughout the distillation process and inference.

All methods were trained and tested on the same fastMRI splits, with 320 × 320 images normalized to [0, 1] and the same undersampling masks (E4×, E8×, P12×, P15×; E: equispaced Cartesian, P: Poisson-disc). Checkpoints were selected by highest validation PSNR for U-Net and ISTA-Net, and by lowest training loss for Score-MRI, DDIM, PD, and ReCoFlow-MRI. ReCoFlow-MRI was evaluated with one-step and five-step Euler integration.

To quantitatively evaluate the reconstruction quality, peak signal-to-noise ratio (PSNR) and structural similarity index (SSIM) were calculated. To assess geometric distortion correction accuracy, 3718 fiducial markers were extracted from the grid phantom to quantify the residual geometric errors.

\subsection{Results on the fastMRI Dataset}
Image reconstruction results on the fastMRI dataset with 1D equispaced Cartesian masks at AFs of $4\times$ and $8\times$ are shown in Fig.~\ref{fig:fastmri_equispaced}. Yellow contours delineate the brain boundaries in the reference images. A mismatch between the yellow reference contours and the brain boundaries in the FT-reconstructed images indicates geometric distortion. In contrast, geometric distortions were corrected in the ZF-corrected images and in images reconstructed by U-Net, ISTA-Net, Score-MRI, DDIM, and ReCoFlow-MRI. ReCoFlow-MRI produced the lowest reconstruction errors among the other methods. Images reconstructed by ReCoFlow-MRI also show better preservation of structural details, as indicated by the red arrows in the zoom-in regions. Reconstruction results using 2D Poisson-disc masks at AFs of $12\times$ and $15\times$ are shown in Fig.~\ref{fig:fastmri_poisson}. Similarly, ReCoFlow-MRI better preserved structural details.

Table~\ref{tab:comparison} summarizes PSNR and SSIM across 1068 fastMRI testing images. Five-step ReCoFlow-MRI yielded the highest PSNR and SSIM across all evaluated undersampling settings. Under $8\times$ equispaced Cartesian undersampling, five-step ReCoFlow-MRI achieved a PSNR of 34.85~dB and an SSIM of 0.930, exceeding the best baseline PSNR and SSIM of 32.63~dB and 0.907, respectively. Similarly, five-step ReCoFlow-MRI improved PSNR by 0.4 and 0.6~dB over the best-performing baseline under $12\times$ and $15\times$ 2D Poisson-disc undersampling, respectively. With a single Euler step, ReCoFlow-MRI exceeded all baselines in PSNR under equispaced Cartesian undersampling and in both PSNR and SSIM under $15\times$ Poisson-disc undersampling.
\begin{figure*}[!t]
\centering
\includegraphics[width=0.92\textwidth]{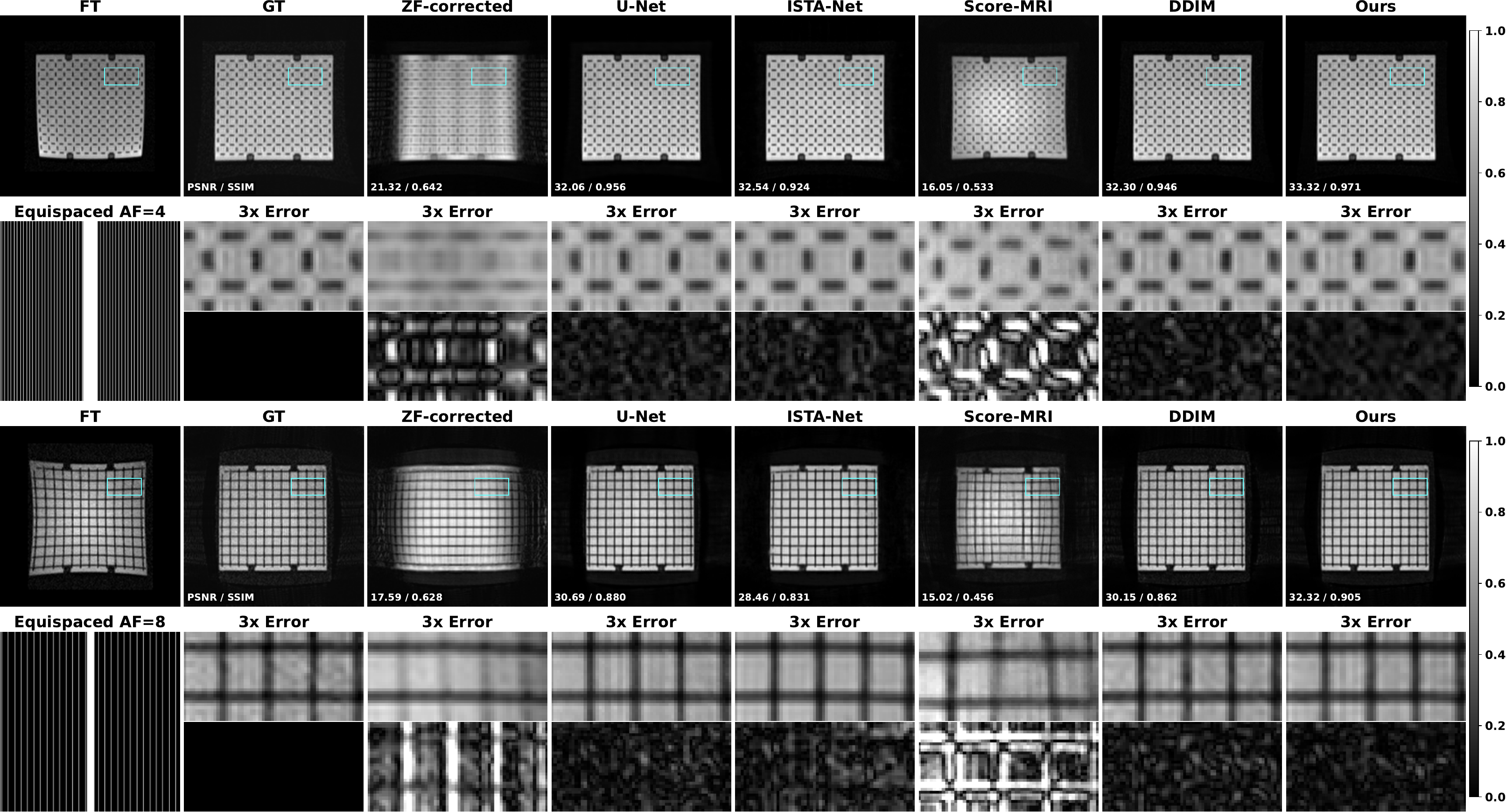}
\caption{Results on grid phantom data with equispaced Cartesian undersampling at acceleration factors of $4\times$ and $8\times$. FT images and fully sampled GT  with undersampled reconstructions by ZF-corrected, U-Net, ISTA-Net, Score-MRI, DDIM, and ReCoFlow-MRI are displayed. The enlarged regions and $3\times$ error maps assess recovery of the grid structure and residual geometric distortion.}
\label{fig:grid_phantom_transfer}
\end{figure*}

\subsection{Results on the Grid Phantom Dataset}
\begin{table}[!htbp]
\caption{Geometric errors at the 3718 fiducial markers before and after GNL correction.}
\label{tab:gnl_marker_error}
\centering
\small
\renewcommand{\arraystretch}{1.15}
\begin{tabular*}{\columnwidth}{@{\extracolsep{\fill}}llcc@{}}
\toprule
Status & Error range & Markers, $n$ (\%) & Max. error (mm) \\
\midrule
Uncorrected & $<5$~mm     & 2173 (58.45) & 24.9515 \\
            & $5$--$10$~mm  & 1287 (34.62) &          \\
            & $10$--$25$~mm & 258 (6.94)   &          \\
\midrule
Corrected   & $<1$~mm       & 3658 (98.39) & 2.1233   \\
            & $1$--$2$~mm   & 59 (1.59)    &          \\
            & $2$--$2.5$~mm & 1 (0.03)     &          \\
\bottomrule
\end{tabular*}
\end{table}
The grid phantom was scanned on the Australian MRI-Linac system to evaluate the reconstruction performance of the proposed method, as shown in Fig.~\ref{fig:grid_phantom_transfer}. Phantom grids become warped in FT-reconstructed images, showing considerable geometric distortions. Less blurring and artifacts are visible on ReCoFlow-reconstructed images in comparison with the other models. The zoom-in regions and error maps demonstrated that ReCoFlow-MRI resulted in better image quality than the existing models under $4\times$ and $8\times$ equispaced Cartesian undersampling.
 
\begin{figure}[!t] 
\centering 
\begin{minipage}[t]{0.49\columnwidth} 
\centering 
\includegraphics[width=\linewidth,trim=70 14 44 8,clip]{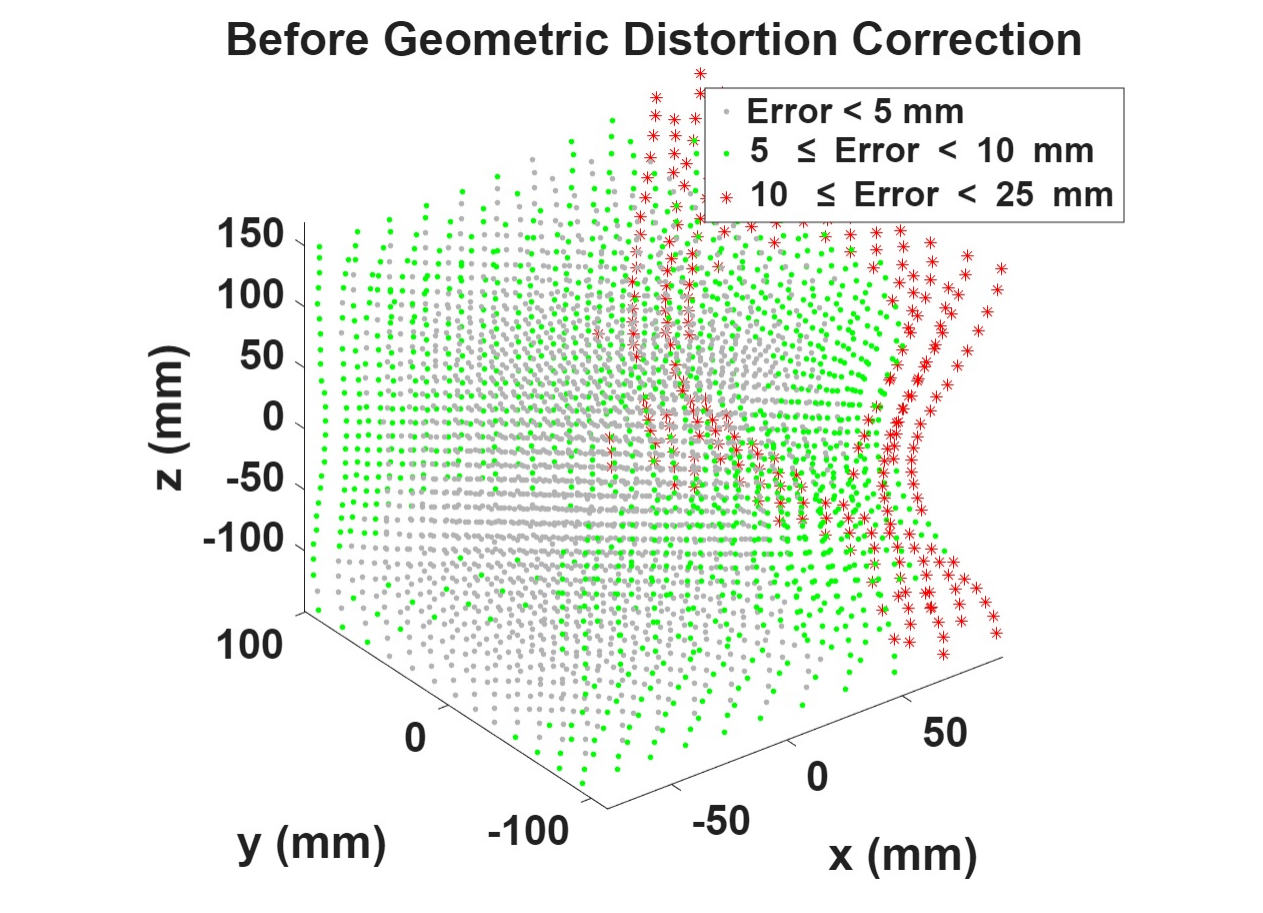}\\[-2pt] 
\textnormal{(a)} 
\end{minipage} 
\hfill
\begin{minipage}[t]{0.49\columnwidth} 
\centering 
\includegraphics[width=\linewidth,trim=70 14 44 8,clip]{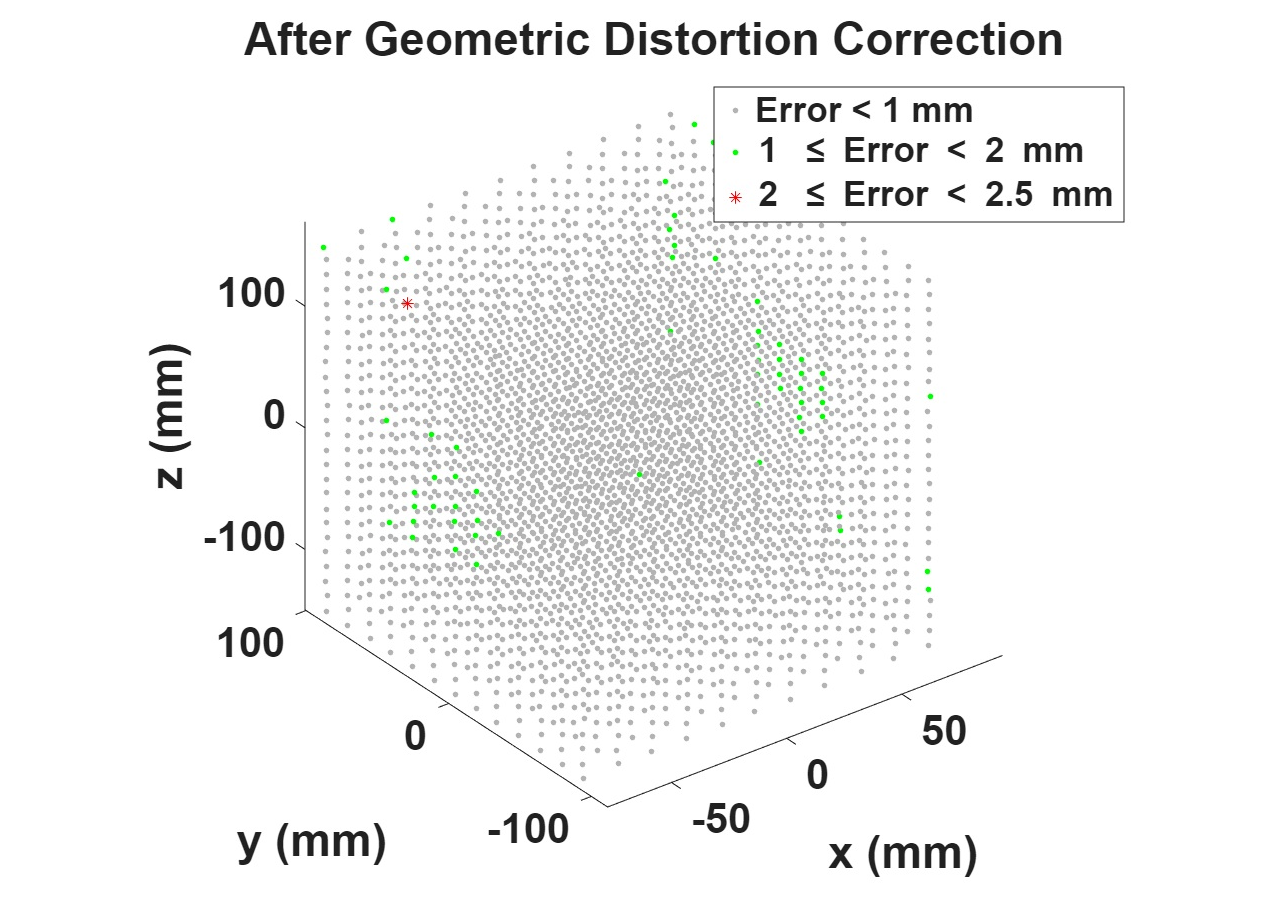}\\[-2pt] 
\textnormal{(b)} 
\end{minipage} 
\caption{Spatial distributions of geometric errors at the 3718 fiducial markers before and after GNL correction. (a) Uncorrected marker positions, categorized as errors below 5~mm, 5--10~mm, and 10--25~mm. (b) Corrected marker positions, categorized as errors below 1~mm, 1--2~mm, and 2--2.5~mm.} 
\label{fig:gnl_marker_error} 
\end{figure}

3718 fiducial markers were extracted from the grid phantom to quantitatively evaluate the geometric distortions. As shown in Table~\ref{tab:gnl_marker_error}, 1545 markers  (41.55\%) had geometric displacement larger than 5~mm before correction. After GNL correction, 3658 markers (98.39\%) had residual distortions below 1~mm and 59 markers (1.59\%) had errors of 1--2~mm. The maximal displacement decreased from 24.95 to 2.12~mm, displaying a 91.5\% reduction. Fig.~\ref{fig:gnl_marker_error} shows the spatial distributions of the fiducial markers, indicating that the proposed method caused dramatic reduction on GNL-induced image distortions.

\begin{figure*}[!t]
\centering
\includegraphics[width=0.92\textwidth]{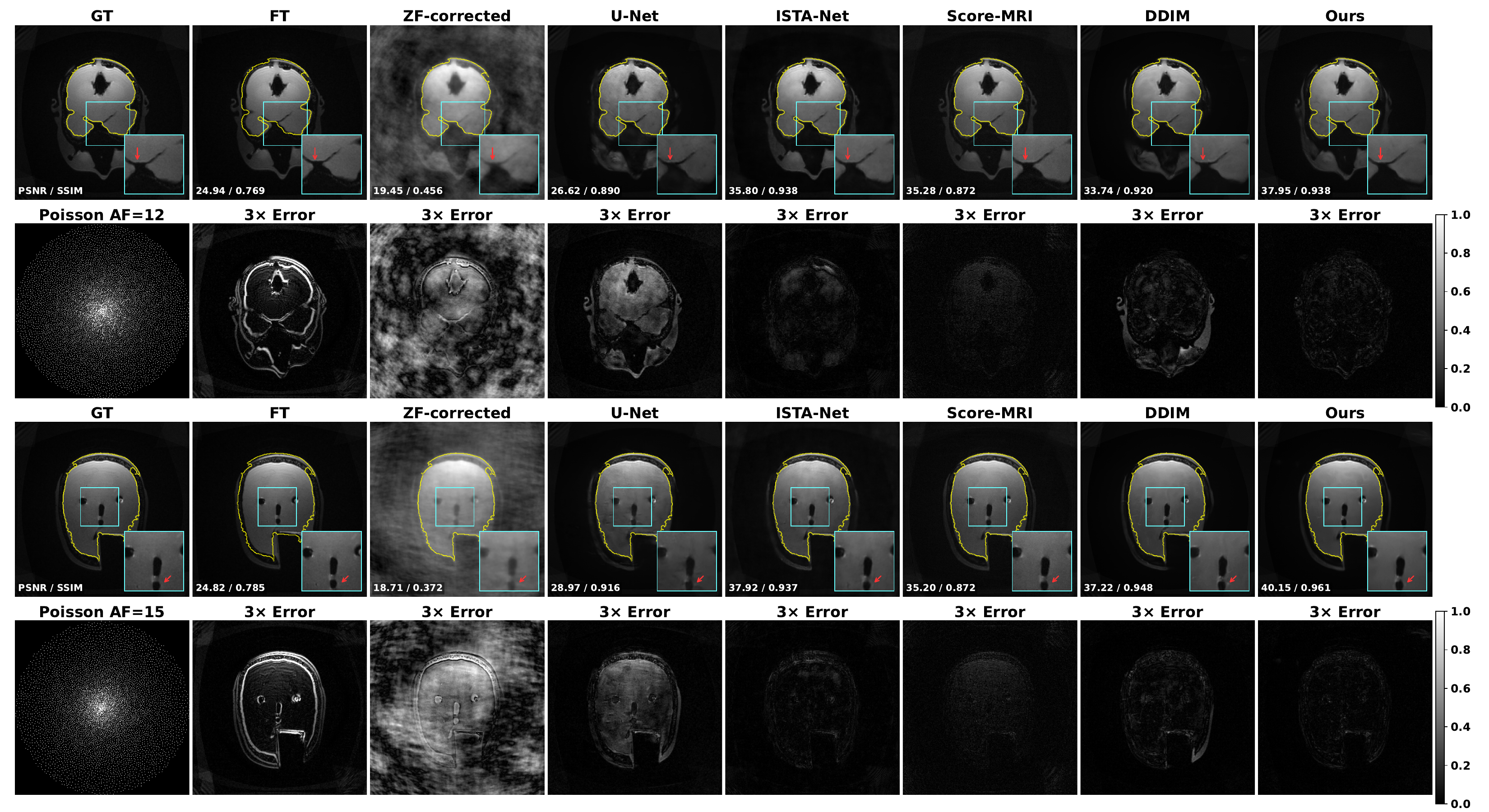}
\caption{Brain phantom images reconstructed with Poisson-disc undersampling at acceleration factors of $12\times$ and $15\times$. Yellow lines represent the brain contour of GT images. Zoom-in areas show local phantom structures and the $3\times$ error maps are displayed at the bottom.}
\label{fig:mrilinac_brain_phantom}
\end{figure*}

\subsection{Results on Brain Phantom and Volunteer Dataset}
\begin{figure*}[!t]
\centering
\includegraphics[width=0.92\textwidth]{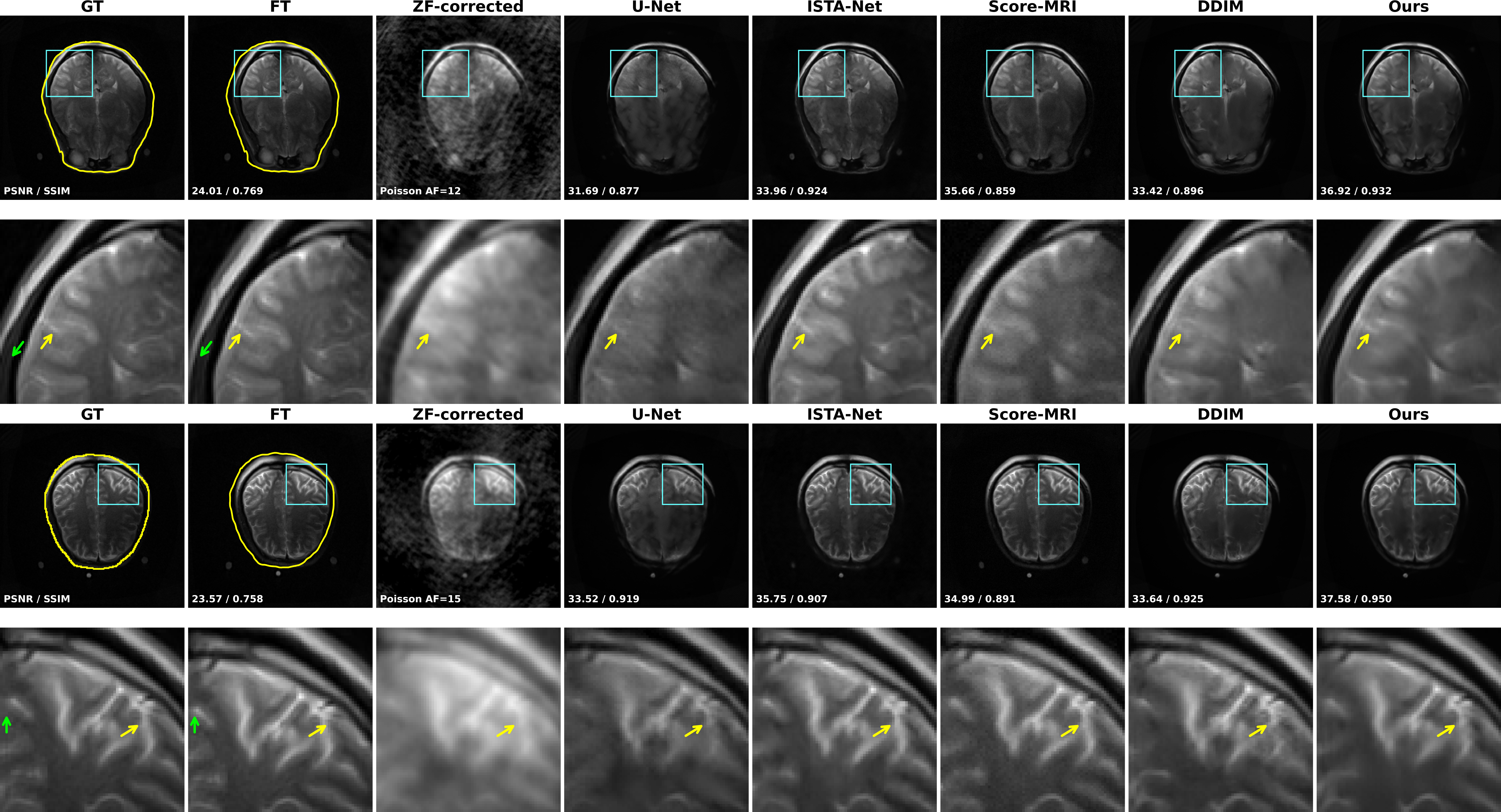}
\caption{Results on volunteer brain images with $12\times$ and $15\times$ Poisson-disc undersampling, respectively. Enlarged regions provide a detailed comparison of geometric correction and anatomical recovery. Green arrows indicate GNL-induced spatial displacement and yellow arrows highlight anatomical structure details.}
\label{fig:mrilinac_volunteer}
\end{figure*}

Images of a brain phantom were acquired on the MRI-Linac scanner to evaluate the performance of ReCoFlow-MRI. As shown in Fig.~\ref{fig:mrilinac_brain_phantom}, yellow lines do not align with brain phantom contours in the FT images because of geometric distortions. Whereas, ReCoFlow-MRI-reconstructed images show better alignment with the yellow lines, demonstrating reduced geometric displacements. Compared with the existing models, ReCoFlow-MRI provided highest PSNR and SSIM values for both $12\times$ and $15\times$ Poisson-disc undersampled acquisitions, showing a similar performance trend to that observed on the fastMRI dataset in Table~\ref{tab:comparison}.
Fig.~\ref{fig:mrilinac_volunteer} shows the volunteer brain reconstruction results with 2D Poisson-disc undersampling. Similar to the results of Fig.~\ref{fig:fastmri_equispaced}  and  Fig.~\ref{fig:fastmri_poisson}, ReCoFlow-MRI preserved finer anatomical structures than the other methods, as indicated by the yellow arrows. It suggests effective distortion correction and high-fidelity image reconstruction, simultaneously. 

\subsection{Ablation Studies}
In this work, ablation studies were designed to evaluate the contributions of the GNL-aware conditioning image, number of ODE integration steps and the pixel-supervised endpoint loss on the fastMRI brain testing dataset.

\begin{table*}[!t]
\caption{Effect of the GNL-Aware conditioning image
$x_{\mathrm{zf}}^{\mathrm{correct}}$ on the fastMRI test set.
Best results are shown in bold.}
\label{tab:ablation}
\centering
\small
\begin{tabular}{|c|c|c|c|c|c|c|c|c|c|}
\hline
\multicolumn{5}{|c|}{fastMRI: Equispaced Cartesian Masks} &
\multicolumn{5}{c|}{fastMRI: Poisson-disc Masks} \\
\hline
Mask & Variant & Steps & PSNR (dB) & SSIM
& Mask & Variant & Steps & PSNR (dB) & SSIM \\
\hline

E4$\times$ & $-x_{\mathrm{zf}}^{\mathrm{correct}}$ & 250 & 32.06 & 0.906
& P12$\times$ & $-x_{\mathrm{zf}}^{\mathrm{correct}}$ & 250 & 31.87 & 0.861 \\
\hline

E4$\times$ & $+x_{\mathrm{zf}}^{\mathrm{correct}}$ & 1 & 38.05 & 0.948
& P12$\times$ & $+x_{\mathrm{zf}}^{\mathrm{correct}}$ & 1 & 35.11 & 0.917 \\
\hline

E4$\times$ & $+x_{\mathrm{zf}}^{\mathrm{correct}}$ & 5 & \textbf{38.60} & \textbf{0.956}
& P12$\times$ & $+x_{\mathrm{zf}}^{\mathrm{correct}}$ & 5 & \textbf{35.35} & \textbf{0.922} \\
\hline

E8$\times$ & $-x_{\mathrm{zf}}^{\mathrm{correct}}$ & 250 & 26.96 & 0.773
& P15$\times$ & $-x_{\mathrm{zf}}^{\mathrm{correct}}$ & 250 & 29.79 & 0.730 \\
\hline

E8$\times$ & $+x_{\mathrm{zf}}^{\mathrm{correct}}$ & 1 & 34.16 & 0.915
& P15$\times$ & $+x_{\mathrm{zf}}^{\mathrm{correct}}$ & 1 & 34.84 & 0.915 \\
\hline

E8$\times$ & $+x_{\mathrm{zf}}^{\mathrm{correct}}$ & 5 & \textbf{34.42} & \textbf{0.924}
& P15$\times$ & $+x_{\mathrm{zf}}^{\mathrm{correct}}$ & 5 & \textbf{34.96} & \textbf{0.919} \\
\hline

\end{tabular}
\end{table*}

\subsubsection{Effect of the GNL-Aware Conditioning Image}
As shown in Table~\ref{tab:ablation}, removing $x_{\mathrm{zf}}^{\mathrm{correct}}$ substantially reduces reconstruction accuracy under all four sampling settings. 250 inference steps were selected for the unconditional model as fewer steps yielded lower reconstruction quality, whereas increasing the step count beyond 250 provided little improvement. Even with 250 sampling steps, the unconditional model produced lower PSNR and SSIM than the conditional model evaluated using five steps. The largest PSNR difference is observed under $8\times$ equispaced Cartesian sampling, where the GNL-aware conditioning image improves PSNR from 26.96 to 34.42~dB. These results show that the acquisition-specific condition is required to improve reconstruction performance and to support few-step inference.

\subsubsection{Effect of the Number of ODE Integration Steps}
\begin{figure}[!t]
\centering
\includegraphics[width=0.92\columnwidth]{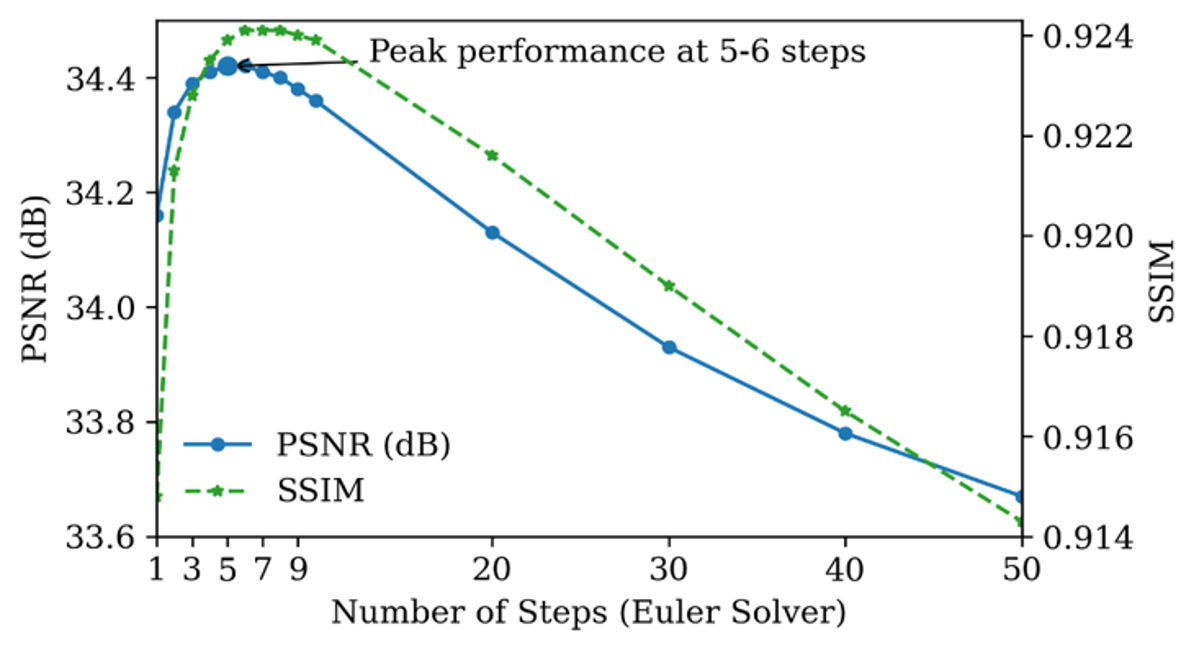}
\caption{Effect of the number of Euler steps under the E8$\times$ mask. The checkpoint used for this analysis was trained with the conditional flow-matching loss without $\mathcal{L}_{\mathrm{pix}}$.}
\label{fig:steps_before_pixel_finetuning}
\end{figure}

 As shown in Fig.~\ref{fig:steps_before_pixel_finetuning}, PSNR increases from one to five Euler steps and reaches its maximum at five steps, whereas SSIM reaches its maximum at six steps. Both values decrease gradually after six steps. Therefore $K=5$ was selected for subsequent endpoint reconstruction and pixel-supervised endpoint fine-tuning in this work.

The pretrained model was then fine-tuned using the pixel-supervised
endpoint loss computed from its five-step reconstruction. Consequently, the fine-tuned checkpoint was evaluated using the same five-step Euler solver.

\subsubsection{Effect of the Pixel-Supervised Endpoint Loss}
Table~\ref{tab:pixel_loss_ablation} compares ReCoFlow-MRI trained with and without the pixel-supervised endpoint loss $\mathcal{L}_{\mathrm{pix}}$. Adding $\mathcal{L}_{\mathrm{pix}}$ consistently improves both PSNR and SSIM under all evaluated masks. The PSNR gains are 0.57, 0.43, 0.43, and 0.21~dB under the E4$\times$, E8$\times$, P12$\times$, and P15$\times$ masks, respectively. These results show that direct endpoint supervision complements conditional velocity matching and improves the accuracy of the final reconstruction.

\begin{table}[!t]
\caption{Effect of the pixel-supervised endpoint loss on the fastMRI test set. Best results are shown in bold.}
\label{tab:pixel_loss_ablation}
\centering
\small
\renewcommand{\arraystretch}{1.15}
\begin{tabular*}{\columnwidth}{@{\extracolsep{\fill}}lcccc@{}}
\toprule
& \multicolumn{2}{c}{Without $\mathcal{L}_{\mathrm{pix}}$}
& \multicolumn{2}{c}{With $\mathcal{L}_{\mathrm{pix}}$} \\
\cmidrule(lr){2-3}\cmidrule(lr){4-5}
Mask & PSNR & SSIM & PSNR & SSIM \\
\midrule
E4$\times$  & 38.60 & 0.956 & \textbf{39.17} & \textbf{0.961} \\
E8$\times$  & 34.42 & 0.924 & \textbf{34.85} & \textbf{0.930} \\
P12$\times$ & 35.35 & 0.922 & \textbf{35.78} & \textbf{0.928} \\
P15$\times$ & 34.96 & 0.919 & \textbf{35.17} & \textbf{0.922} \\
\bottomrule
\end{tabular*}
\end{table}

\section{Discussion}
\subsection{Generalization and Adaptability of ReCoFlow-MRI}
The generalization ability of MRI reconstruction models is important for their application to different imaging conditions and anatomical structures. Although ReCoFlow-MRI was initially trained on the fastMRI brain dataset, the pretrained model was directly applied to brain phantom and in vivo volunteer brain data acquired on the MRI-Linac system without additional training or fine-tuning. ReCoFlow-MRI provided high-quality reconstructions with reduced undersampling artifacts and improved anatomical fidelity compared with the evaluated baseline methods (see  Fig.~\ref{fig:mrilinac_brain_phantom} and Fig.~\ref{fig:mrilinac_volunteer}). It demonstrates the generation ability for previously unseen MRI-Linac acquisitions, indicating its potential applicability across different imaging conditions.

For the grid phantom, transfer learning was employed to adapt the pretrained ReCoFlow-MRI model and the other baseline models. A total of 302 grid-phantom images were acquired on the MRI-Linac system, of which 20 uniformly spaced images were reserved as a testing dataset and the remaining 282 images were used for fine-tuning. All the models were initialized using the checkpoint pretrained on the fastMRI brain dataset and subsequently fine-tuned on the grid-phantom training images. As shown in Fig.~\ref{fig:grid_phantom_transfer}, the fine-tuned ReCoFlow-MRI model effectively suppressed undersampling artifacts and recovered grid structures, demonstrating its adaptability to grid-phantom reconstruction through transfer learning.

\subsection{Selection of Inference Steps and Comparison With Progressive Distillation}

The reconstruction performance of ReCoFlow-MRI largely depends on the number of numerical integration steps. Imaging results have shown that the conditional flow-matching model achieves its highest PSNR at five Euler steps, while SSIM reaches its maximum at six steps (see Fig.~\ref{fig:steps_before_pixel_finetuning}). Increasing the number of steps beyond these values does not further improve reconstruction accuracy. It suggests that more numerical integration of the learned velocity field may add details from the initial noise rather than from the conditioning image $c$. Therefore, five-step inference was selected for subsequent endpoint fine-tuning here. 

It is also noted that single-step ReCoFlow-MRI without pixel-supervised endpoint fine-tuning outperforms all evaluated baseline methods in terms of PSNR for equispaced Cartesian sampling. For Poisson-disc sampling, it outperforms all baselines at 15$\times$ acceleration, while remaining slightly below ISTA-Net at 12$\times$ acceleration, demonstrating superior reconstruction ability of the conditional flow-matching model. 

PD reduces inference steps by transferring the sampling behavior of a pretrained model to successive student networks. To investigate its effect on reconstruction accuracy, we evaluated the 128-step DDIM teacher and student models which were progressively distilled to 64, 32, 16, and 8 steps on the fastMRI testing dataset. As shown in Table~\ref{tab:pd_distillation}, the first distillation stage largely preserved the teacher's reconstruction performance, with slight improvements under E4$\times$ and P15$\times$ accelerations. However, further reducing the sampling steps from 64 to 8 lead to a consistent decline in PSNR across all four undersampling masks. Compared with the 128-step teacher, the 8-step student exhibits PSNR reductions ranging from 0.08 to 1.05~dB, with the largest degradation observed under $8\times$ equispaced Cartesian sampling. It indicated the trade-off between sampling efficiency and reconstruction accuracy in the PD-based model. In comparison, the five-step ReCoFlow-MRI model had higher PSNR and SSIM than the 128-step DDIM teacher and all evaluated PD students (64, 32, 16, and 8 steps) across all four undersampling settings. 

\begin{table}[!t]
\caption{Progressive-distillation results of the 128-step DDIM teacher and its distilled student models on the fastMRI test set. Each entry reports PSNR/SSIM; best results are shown in bold.}
\label{tab:pd_distillation}
\centering
\small
\renewcommand{\arraystretch}{1.15}
\setlength{\tabcolsep}{4pt}
\resizebox{\columnwidth}{!}{%
\begin{tabular}{lcccc}
\toprule
Model (steps) & E4$\times$ & E8$\times$ & P12$\times$ & P15$\times$ \\
\midrule
DDIM (128) & 36.84/0.950 & \textbf{32.63}/0.907 & \textbf{33.30}/0.903 & 32.85/0.898 \\
PD (64)  & \textbf{36.89/0.952} & 32.34/\textbf{0.908} & 33.23/\textbf{0.904} & \textbf{32.89/0.900} \\
PD (32)  & 36.83/\textbf{0.952} & 32.08/0.906 & 32.99/0.901 & 32.72/0.899 \\
PD (16)  & 36.79/0.951 & 31.87/0.903 & 32.82/0.898 & 32.54/0.897 \\
PD (8)   & 36.76/0.950 & 31.58/0.897 & 32.64/0.897 & 32.33/0.895 \\
\bottomrule
\end{tabular}%
}
\end{table}

\subsection{Limitations and Future Work}
In this work, grid phantom, brain phantom and volunteer brain images were acquired on an open-bore MRI-Linac scanner with retrospective undersampling masks to evaluate the performance of the ReCoFlow-MRI. Prospectively undersampling acquisitions pose additional challenges for deep-learning based MRI reconstructions as undesired noise is normally presented in the acquired k-space data. The performance of ReCoFlow-MRI model on prospectively undersampled k-space data requires further investigation. In addition to GNL, $B_0$ inhomogeneity in static magnetic $B_0$ field can also cause geometric distortions~\cite{Shan2024}. However, $B_0$ inhomogeneity-induced distortions change with sequence parameters such as gradient strength, gradient polarity and imaging bandwidth. In the future, we will incorporate $B_0$ inhomogeneity information into ReCoFlow to account for $B_0$ inhomogeneity distortions.

\section{Conclusion}
In this study, we proposed ReCoFlow-MRI, a conditional flow-matching framework for GNL-induced nonlinear MRI reconstruction. By incorporating calibrated nonlinear spatial encoding into the conditioning image, the proposed method jointly addresses undersampling artifacts and GNL-induced geometric distortions within a few-step generative reconstruction process. Extensive experiments on fastMRI, MRI-Linac phantom and volunteer images demonstrate superior reconstruction performance over U-Net, ISTA-Net, and diffusion-based methods across multiple undersampling settings. ReCoFlow-MRI shows promises for high-fidelity and distortion-corrected MRI reconstruction, and will facilitate applications in accurate MRI-guided radiotherapy.

\bibliographystyle{IEEEtran}
\bibliography{mybib}

\end{document}